\documentclass[12pt]{article}
\usepackage{graphicx}
\usepackage{amssymb}
\usepackage{epsfig}
\usepackage{amsmath}
\newcommand{\be}{\begin{equation}}
\newcommand{\ee}{\end{equation}}

\newcommand{\bea}{\begin{eqnarray}}
\newcommand{\eea}{\end{eqnarray}}

\newcommand{\p}{\partial}
\newcommand{\half}{\frac{1}{2}}
\newcommand{\m}{\mu}
\newcommand{\n}{\nu}

\newcommand{\g}{\gamma}

\newcommand{\ta}{\theta}
\newcommand{\as}{AdS_5 \times S^5}

\DeclareSymbolFont{AMSa}{U}{msa}{m}{n}
\DeclareSymbolFont{AMSb}{U}{msb}{m}{n}
\DeclareMathSymbol{\fieldR}{\mathalpha}{AMSb}{"52}

\begin{document}
\begin{titlepage}
\begin{center}
\hfill {\tt YITP-SB-05-44}\\
\hfill {\tt hep-th/0512101}\\
\vskip 10mm

{\Large {\bf Half-BPS Geometries and \\ \vskip 0.3cm Thermodynamics of Free Fermions}}

\vskip 10mm
\renewcommand{\thefootnote}{\fnsymbol{footnote}}

{\bf Simone Giombi, Manuela Kulaxizi,\\ Riccardo Ricci and Diego
Trancanelli \footnote{E-mails: {\tt 
sgiombi, kulaxizi, rricci, dtrancan@insti.physics.sunysb.edu}}}

\vskip 4mm {\em C. N. Yang Institute for Theoretical
Physics}\\
{\em State University of New York at Stony Brook}\\
{\em Stony Brook, NY 11794-3840, USA}
\end{center}
\vskip 0.8cm

\renewcommand{\thefootnote}{\arabic{footnote}}
\setcounter{footnote}{0}

\begin{center} {\bf ABSTRACT }\end{center}
\begin{quotation}
\noindent
Solutions of type IIB supergravity which preserve half of the supersymmetries
have a dual description in terms of free fermions, as elucidated by the ``bubbling AdS" construction of Lin, Lunin and Maldacena.
In this paper we study the half-BPS geometry associated 
with a gas of free fermions in thermodynamic equilibrium obeying the Fermi-Dirac distribution.
We consider both regimes of low and high temperature. In the former case,
we present a detailed computation of the ADM mass of the supergravity solution and 
find agreement with the thermal energy of the
fermions. 
The solution has a naked null singularity and, by general arguments, is expected to develop a finite area horizon once stringy corrections are included.
By introducing a stretched horizon, we propose a way to match the entropy of the fermions with the
entropy of the geometry in the low temperature regime.
In the opposite limit of high temperature, the solution 
resembles a dilute gas of D3 branes. 
Also in this case the ADM mass of the geometry agrees with the thermal energy of the fermions.

\end{quotation}
\vfill \flushleft{December 2005}
\end{titlepage}
\eject

%%%%%%%%%%%%%%%%%%%%%%%%%

\newpage
\tableofcontents

%%%%%%%%%%%%%%%%%%%%%%%%%%
\section{Introduction}
%%%%%%%%%%%%%%%%%%%%%%%%%%

According to the AdS/CFT correspondence, deformations of AdS
geometries should map to states in the dual CFT living at the
boundary of AdS \cite{Maldacena:1997re}. Recently a concrete
realization of this map has
been found for the important sector of half-BPS operators of
${\cal N}=4$ Super Yang-Mills. These operators have conformal
dimension $\Delta$ equal to the $U(1)_R$ charge. They form a
decoupled sector of ${\cal N}=4$ Super Yang-Mills which can be
efficiently described by a gauged quantum mechanics matrix model
with harmonic oscillator potential, see \cite{Berenstein:2004kk}
and for earlier work \cite{Corley:2001zk}.
The matrix model is well known to be completely integrable.
The main reason behind integrability
is that, in the eigenvalue basis, the eigenvalues behave as
fermions in a harmonic potential. In the semiclassical limit the
half-BPS states can be depicted as droplets of fermions in a
two-dimensional phase space. One expects then the following
AdS/CFT dictionary. Small ripples above the Fermi sea correspond
to graviton excitations of $\as$. Small holes below the Fermi energy
correspond to giant gravitons, while small droplets of fermions
outside the Fermi sea map to dual giant gravitons.
All this is very reminiscent of both old
and recent works on $c=1$ string theory and its matrix model
reinterpretation \cite{Klebanov:1991qa} \cite{McGreevy:2003kb} (for a recent
review see \cite{Nakayama:2004vk}).

Remarkably this whole picture has found an impressive confirmation
through the explicit construction of the full moduli space of
half-BPS IIB supergravity solutions discovered by Lin, Lunin and
Maldacena (LLM) \cite{Lin:2004nb} \footnote{For related work see \cite{all}.}. 
The phase space distribution of
the matrix model eigenvalues is in one-to-one correspondence with
IIB supergravity backgrounds which preserve half of the
supersymmetry. Moreover the two-dimensional phase space of the
fermions has an interesting physical embedding in the space-time
geometry. At the quantum level the incompressibility of the droplets
in phase space (due to Fermi-Dirac statistics) corresponds in the
dual supergravity side to the requirement that the Ramond-Ramond
five-form flux is quantized. The whole family of half-BPS geometries
can be constructed in terms of an auxiliary function $z$ which also
determines the fermion distribution. The regularity of the
supergravity background amounts to requiring a suitable boundary
condition on the auxiliary function. The  AdS ``bubbling
geometries'' are therefore in general smooth supergravity
backgrounds.

The fermions discussed so far are characterized by having a
step-function distribution in the two-dimensional phase space.
They can be seen therefore as fermions at zero ``temperature''. It
is then natural to investigate how turning on the temperature
affects the supergravity solution. The fermion at non zero
temperature are described by a Fermi-Dirac distribution. The
corresponding AdS ``bubbling'' solution has been first obtained
in \cite{Buchel:2004mc} and further studied in \cite{Balasubramanian:2005mg}
where it was given the name of {\it hyperstar}.  This supergravity background
can be thought of as
resulting from a coarse graining process of smooth half-BPS
geometries. The fermion distribution of the hyperstar fails to
satisfy the boundary conditions necessary to obtain a smooth
gravity solution. Quite generally when the smoothness condition is
not satisfied naked singularities occur \cite{Caldarelli:2004mz} \cite{Milanesi:2005}.
One expects that $\alpha'$ string corrections will modify the
geometry in proximity of the singularity and that a horizon will
be generated \cite{Dabholkar:2004yr} \cite{Lunin:2002qf}.
This class of singular supergravity solution can
therefore be regarded as incipient black holes. An example of singular
LLM solution is the \emph{superstar} \cite{Myers:2001aq}, which has been
investigated from this point of view  in \cite{Suryanarayana:2004ig} \cite{Shepard:2005zc}.

The duality between fermion distributions and supergravity
solutions at zero temperature suggests that the thermodynamic
properties of the fermion gas at finite temperature should agree
with the corresponding quantities in the supergravity side. In
particular, one expects agreement between the thermal excitation
energy of the fermions and the ADM mass of the supergravity
solution. We will check that this is indeed the case in the two
opposite regimes of low and high temperature.

As we have already remarked, the hyperstar geometry is singular. The singularity
is resolved quantum mechanically through the appearance of a finite
area horizon. One can then use the Bekenstein-Hawking formula to
compute the associated entropy. By placing a stretched horizon in
the hyperstar geometry we propose a way to match the supergravity entropy
with the thermal entropy of the fermions in the low temperature regime, 
up to a numerical factor.

Similarly we investigate the opposite regime of high
temperature. In this limit the Fermi-Dirac distribution reduces to
the classical Boltzmann distribution. We find that in this regime
the metric is reminiscent of the so called dilute gas limit of LLM
configurations associated to the Coulomb branch of Super
Yang-Mills.

The organization of the paper is the following. In the next
section we present a concise description of the main results of
LLM. In sec. 3 we review the thermodynamics of 1D fermions in a
harmonic potential. In sec. 4 we introduce the hyperstar geometry,
discuss its properties in the low temperature regime, compute the
ADM energy, angular momentum and entropy and we compare the results to the
fermion prediction. In sec. 5 we move on to consider the high
temperature limit of the hyperstar distribution, we find the
corresponding metric and determine its mass and angular momentum. We conclude discussing some open questions in sec. 6.

%%%%%%%%%%%%%%%%%%%%%%%%%%%%%%%%%%%%%%
\section{Review of LLM}
%%%%%%%%%%%%%%%%%%%%%%%%%%%%%%%%%%%%%

In this paragraph we briefly review the LLM construction \cite{Lin:2004nb} of $1/2$
BPS IIB supergravity backgrounds. These solutions correspond, in the dual CFT side,
to states satisfying the BPS condition \be \Delta=J \ee where
$\Delta$ is the corresponding conformal dimension and $J$ is a particular
$U(1)$ charge of the $SO(6)$ R-symmetry group.
 By selecting one generator of the $SO(6)$
R-symmetry group of ${\cal N}=4$ Super Yang-Mills we obtain a theory with $SO(4)\times
SO(4)\times R$ bosonic symmetry. In the dual supergravity description
 we look therefore for solutions with this isometry group. 
Assuming that axion and dilaton
 are constant and that only the self-dual five-form 
field strength is turned on, the Ansatz for
the background is
 \begin{eqnarray}
ds^2&=&g_{\mu\nu}dx^\mu dx^\nu +e^{H+G}d\Omega_3^2+e^{H-G}d{\tilde\Omega}_3^2\\
F_{(5)}&=&F_{\mu\nu}dx^\mu\wedge dx^\nu\wedge d\Omega_3+ {\tilde
F}_{\mu\nu}dx^\mu\wedge dx^\nu\wedge d{\tilde\Omega}_3
 \end{eqnarray}
where the Greek indices $\mu,\nu$ run over $0,\ldots,3$. The two
three-spheres $S^3$ and $\tilde S^3$ in the metric make the
$SO(4)\times SO(4)$ isometries manifest. The additional $R$
isometry
 corresponds to the Hamiltonian $\Delta-J$.

 For a background to be  half-BPS 
there should exist a solution to the Killing spinor equation.
Analyzing this equation, LLM were able to prove that the generic
1/2 BPS IIB supergravity background takes the form
  \begin{eqnarray} ds^2 &=& - h^{-2}
(dt + V_i dx^i)^2 + h^2 (dy^2 + dx^idx^i) + y e^{G
 } d\Omega_3^2 + y e^{ - G} d \tilde \Omega_3^2
\\
h^{-2} &=& 2 y \cosh G ,
\\
 y \partial_y V_i &=& \epsilon_{ij} \partial_j z,\;\;\;\;\;\;\;\;\;\;
 y (\partial_i V_j-\partial_j V_i) = \epsilon_{ij} \partial_y z
 \label{diffeqplane}\\
 z &=&\frac{1}{2} \tanh G
 \\
F &=& dB_t \wedge (dt + V) + B_t dV + d \hat B ~,~~~~~~ \nonumber
\\
 \tilde F &=&
d\tilde B_t \wedge (dt + V) + \tilde B_t dV + d \hat { \tilde B}
 \\
 B_t &=& - \frac{1}{4} y^2 e^{2 G
 } ,~~~~~~~~~~~~~~~~~~\tilde B_t = - \frac{1}{4}  y^2 e^{- 2 G}
\\
 d \hat B &=&  - \frac{1}{4} y^3 *_3 d \left( \frac{ z + \half} {y^2 }\right) ~,~~~~~~~~
d \hat {\tilde B} = - \frac{1}{4} y^3 *_3 d \left( \frac{ z - \half
}{y^2 }\right)
 \end{eqnarray}
where $i=1,2$ and $\star_3$ is the Hodge dual operator for the
flat three-dimensional space parameterized by $x_1,x_2,y$.
 Remarkably, the solution is completely specified in terms of a single
auxiliary function $z(x_1,x_2,y)$ which satisfies the linear
differential equation
\begin{equation}
 \partial_i \partial_i z + y \partial_y \left(\frac{ \partial_y z}{ y}\right)
 =0\label{fund}.
 \end{equation}
 It is important to
note that at $y$ equals zero the product of the radii of the two
privileged three-spheres is zero. Therefore, to avoid singular
geometries, the auxiliary function  $z$ must satisfy a suitable
boundary condition. This smoothness condition turns out to be $z=\pm
\half$ on the boundary plane $y=0$. In the limit $z\rightarrow 1/2$
the $\tilde S^3$ sphere shrinks to zero while the other three-sphere
remains finite. The reverse statement applies when $z\rightarrow
-1/2$. It is conventional to assign black and white colors
respectively to the $z=-1/2$ and $z=1/2$ points in the $(x_1,x_2)$
plane. If $\cal{D}$ denotes a black region in this plane, the
energy of the associated supergravity solution has the simple
expression \be
 \Delta=J=\int_{\cal{D}}\frac{d^2x}{2\pi \hbar}\ \half \frac{(x_1^2+x_2^2)}
{\hbar}-\half \left(\int_{\cal{D}}\frac{d^2x}{2\pi h}\right)^2\,.
\ee  The ${\mathbb R}^2$ plane has then a natural interpretation as
the phase space of one-dimensional fermions in a harmonic potential.
This nicely matches the matrix model description in the dual CFT
side \cite{Berenstein:2004kk}. It emerges a beautiful picture
of the moduli space of half-BPS geometries of IIB supergravity in
terms of configurations of droplets of fermions on the $(x_1,x_2)$
plane.
 Note that the fundamental equation (\ref{fund}) has the symmetry $z\rightarrow -z$ which
  simply exchanges the
$S^3$ and $\tilde S^3$ in the solution. In a field theory
description of the fermions, this symmetry amounts to a
particle-hole duality.

The quantization condition on the total area $\cal{A}$ of the
droplets is related to the  five-form flux $N$ as follows \be \frac{{\cal
A}}{2\pi\hbar}=N \ee with \be \hbar=2\pi l_p^4. \ee
The flux $N$ coincides with the number of fermions.
The
simplest configuration in phase space is a black circular droplet
of radius $R_0=\sqrt{2\hbar N}$ and the associated geometry is $AdS_5\times
S^5$ with $N$ units of the five-form flux.
This background has $\Delta=J=0$ and corresponds to the fermion ground state. The boundary of the
droplet can be thought of as the Fermi level of the fermions. The
$S^5$ of the background is obtained by fibering the $\tilde S^3$
sphere on a two-dimensional surface $\Sigma_2$ in the $(x_1,x_2,y)$
space which encircles the droplet. One can easily obtain
configurations with an arbitrary number of $S^5$'s by adding other
droplets. If we deform the circular droplet to
 configurations with different shapes but same area, we obtain backgrounds
with $AdS_5\times S^5$ asymptotics.

The fundamental equation (\ref{fund}) can be rewritten as a
Laplace equation for the quantity $\Phi=z/y^2$ in a six-dimensional space with spherical symmetry in four of the
coordinates. The coordinate $y$ corresponds to the radial
direction in the four-dimensional subspace. This observation
reduces the task of finding the full solution $z(x_1,x_2,y)$ of
eq. (\ref{fund}) to a well known initial-value problem. Once the
boundary condition $z(x_1,x_2,0)$ on the $y=0$ plane is specified,
the solution is
 \be
z(x_1,x_2,y)=\frac{y^2}{\pi}\int_{\mathbb{R}^2}
\frac{z(x_1',x_2',0)dx_1'dx_2'}{[({\bf{x-x'}})^2+y^2]^2}.
\label{zxx}\ee
We can similarly get
 \be
V_i(x_1,x_2,y)=\frac{\epsilon_{ij}}{\pi}
\int_{\mathbb{R}^2}\frac{z(x_1',x_2',0)(x_j-x_j')dx_1'dx_2'}{
[({\bf{x-x'}})^2+y^2]^2}
%={\epsilon_{ij}\over\pi}\int_{\cal{D}}{(x_j-x_j')dx_1'dx_2'\over
%[({\bf{x-x'}})^2+y^2]^2}
.\label{Vxx}\ee

Since we are going to consider only droplet configurations with radial symmetry, it will 
be convenient to rewrite the above formulas in polar coordinates $(x_1,x_2) \rightarrow (R,\phi)$.  It is easy to see that in this case $V_R  =
V_1\cos\phi+V_2\sin\phi=0$. Defining $V \equiv
V_{\phi}  = R (-V_1 \sin\phi+V_2\cos\phi)$ the differential 
equations relating $z$ and $V$
(\ref{diffeqplane}) read  
\bea y\,\p_y\, V\,  =
\, -R\,\p_R\, z \,,~~~~~~~~~~~ \frac{1}{R}\,\p_R\, V\, =\,
\frac{1}{y}\,\p_y\, z . 
\label{diffeqpol}
\eea  
Rewriting eq. (\ref{zxx}) and eq. (\ref{Vxx}) in polar coordinates yields
\bea z(R,y) =   -\int z(R',0)\frac{\p}{\p R'}z_0(R,y;R')dR'
\label{z}
\eea   
\bea
V(R,y) =  \int z(R',0) g_V(R,y;R')dR'
\label{301}
\eea 
where 
\bea z_0(R,y;R') &=&
\frac{R^2-R'^2+y^2}{2\left[(R^2+R'^2+y^2)^2-4R^2R'^2\right]^{1/2}}\\
g_V(R,y;R') &=&
\frac{-2R^2R'(R^2-R'^2+y^2)}{\left[(R^2+R'^2+y^2)^2-4R^2R'^2\right]^{3/2}} . \label{302}
\eea 
We remark that $z_0$ is the LLM function corresponding to
a circular droplet. Indeed in this case 
 $z(R',0) = 1/2 \, \mbox{Sign}(R'-R_0)$ and using eq. (\ref{z}) one obtains $z(R,y) = z_0(R,y;R_0)$. As previously anticipated such a configuration gives rise to the $\as$ solution. In fact performing the following change of coordinates \cite{Lin:2004nb}
\bea y &=& R_0 \sinh \rho \sin \theta \\
R &= & R_0 \cosh\rho \cos \theta\\
\phi &= & \tilde\phi + t \eea 
one recovers the $\as$ metric in standard form 
\be 
ds^2
= R_0 \big( - \cosh^2 \rho d t^2 + d\rho^2+ \sinh^2 \rho d\Omega_3^2
+ d \theta^2 + \cos^2 \theta d \tilde \phi^2 + \sin^2 \theta d\tilde
\Omega_3^2 \big) .
\ee

A variation of the method described so far can be similarly applied
to obtain 1/2 BPS M-theory backgrounds with $AdS_{4,7}\times
S^{7,4}$ asymptotics \cite{Lin:2004nb}. In this case the geometry is
in one-to-one correspondence with solutions of a three-dimensional
Toda equation, which plays the same role as eq. (\ref{fund}).

%%%%%%%%%%%%%%%%%%%%%%%%%%%%%%%%%%%%%%
\section{1D fermions in the harmonic well}
%%%%%%%%%%%%%%%%%%%%%%%%%%%%%%%%%%%%%

In this section we review the basics of the thermodynamics of one-dimensional fermions
in a harmonic potential. In what follows, we will consistently adopt units in which $\hbar=k_B=1$. We consider a gas of $N$ non-interacting fermions with hamiltonian
\bea
H(p,q) = \half (p^2 + q^2)
\eea
in thermodynamic equilibrium at a given temperature $T$. For large $N$, we adopt the semi-classical approximation in which the energy is taken to be a continuous variable.
The probability distribution as a function of the energy $H(p,q)=\epsilon$ is given by the Fermi-Dirac distribution:
\bea
n_{FD}(\epsilon) = \frac{1}{e^{(\epsilon-\mu)/T}+1} 
\label{FD}
\eea
where $\mu$ is the Fermi energy. This is determined by the normalization condition
\bea
\int_0^{\infty} \frac{d \epsilon}{e^{(\epsilon-\mu)/T}+1} = N
\eea
which gives
\bea
\mu = T \ln (e^{N/T}-1).
\label{fermilev}
\eea
We will first consider the limit of very small temperature $T$, or more precisely $N/T \gg 1$. In this limit, 
the Fermi level becomes
\bea
\mu = N + \mathcal{O} (T e^{-N/T}).
\label{fermilevTsmall}
\eea
The total energy of the Fermi gas is given by:
\bea
E = \int_0^{\infty} \frac{\epsilon \, d\epsilon}{e^{(\epsilon-\mu)/T}+1}
\label{totenergy}
\eea
which for small $T$ can be evaluated by means of the Sommerfeld expansion \cite{Landau}
\bea
I=\int_0^{\infty}\frac{f(\epsilon)d\epsilon}{e^{(\epsilon-\m)/T}+1}=
\int_0^{\m}f(\epsilon)d\epsilon + \frac{\pi^2}{6} T^2 f'(\m)
+\frac{7 \pi^4}{360} T^4 f'''(\m)+
\mathcal{O}(T^6).
\label{sommerfeld}
\eea
This gives:
\bea
E \simeq \frac{N^2}{2} + \frac{\pi^2}{6} T^2.
\label{smallTen}
\eea
The first term is clearly the ground state energy of the $N$ fermions, so we expect the dual gravity solution to have a mass (and angular momentum) difference of $\Delta = \frac{\pi^2}{6} T^2$ with respect to the $\as$ background. It is worth noting that in eq. (\ref{smallTen}) we only neglect exponentially suppressed terms. In fact, since $f(\epsilon)=\epsilon$, there
are no power series corrections to the energy beyond $T^2$. This is a specific feature of the 1D harmonic oscillator, and we will recover it in the  energy \footnote{Modulo a subtlety involving $T^4$ terms 
to be discussed later.} and angular momentum of the hyperstar in the low $T$ limit.

To evaluate the entropy of the fermion gas, it is convenient to first obtain the free-energy $F$ of the system. 
This is computed from the partition function $Z$, which in the continuous limit we are considering reads
\bea
Z = \mbox{exp} \left[-{N \mu \over T} + \int_0^{\infty} d\epsilon \, \ln (1+ e^{-(\epsilon-\mu)/T})\right]. 
\eea 
One can verify that this expression for the partition function is correct by checking that the 
relation $E = -{\partial \over \partial \beta} \ln Z$ (where $\beta = 1/T$) is satisfied. 
Using the definition $F=-T \ln Z$ one obtains the free energy 
\bea
F &=& N \mu  - T \int_0^{\infty} d\epsilon \, \ln (1+ e^{-(\epsilon-\mu)/T}) \cr
  &=& N \mu - \int_0^{\infty} d\epsilon \, {\epsilon \over e^{(\epsilon-\mu)/T}+1} = N\mu - E
\eea
where in the second line an integration by parts was made.
The entropy is then given by the relation $F = E - TS$ from which we get
\bea
S = \frac{2E-N \mu}{T}.
\label{entro}
\eea
For small $T$, using eq. (\ref{fermilevTsmall}) and eq. (\ref{smallTen}) one gets
\bea
S \simeq \frac{\pi^2}{3} T
\label{entrosmall}
\eea
where again only exponetially small terms are neglected.

We now consider the opposite limit of very high temperature $N/T \ll 1$. In this limit, the
Fermi distribution clearly reduces to the Boltzmann density
\bea
n_{FD}(\epsilon) \rightarrow n_B(\epsilon) = N \beta \, e^{-\beta \epsilon}
\eea
where $\beta = 1/T$. The total energy in this approximation is readily computed
\bea
E = N T.
\eea
The entropy can be obtained from eq. (\ref{entro}) (which is valid for any temperature) using the large $T$ approximation
\bea
\mu \simeq T \ln N/T
\eea
and reads
\bea
S \simeq N \ln T + 2N-N \ln N.
\eea

%%%%%%%%%%%%%%%%%%%%%%%%%%%%%%%%%%%%%%
\section{The hyperstar: Low temperature regime}
%%%%%%%%%%%%%%%%%%%%%%%%%%%%%%%%%%%%%

We now introduce the half-BPS geometry dual to the Fermi-Dirac gas described 
in the previous section \cite{Buchel:2004mc}. This solution was named hyperstar in 
\cite{Balasubramanian:2005mg}.

A given $z(R,0)$
corresponds to a fermion density $n(R)$ in the phase space via the
relation \bea z(R,0)=\frac{1}{2}-n(R)\label{303} . \eea For example,
the $\as$ solution is associated to the step function density $n_0 =
 \vartheta (R-R_0) $, which can be viewed as the zero temperature
limit of the Fermi-Dirac distribution (\ref{FD}). One can turn on
the temperature on the fermion side by replacing $n_0$ with
$n_{FD}(R)$
 and construct the corresponding supergravity 
background  by using (\ref{z}), (\ref{301}) and (\ref{303}). It is important to remark that the temperature we are turning on is the temperature in the ``auxiliary" description of the free fermion gas. It is not a temperature of the supergravity solution or of the dual gauge theory. Indeed, we remain in the supersymmetric half-BPS sector. It would be interesting to understand better what corresponds to this temperature on the gravity and gauge theory side. For the time being, we regard $T$ just as a deformation parameter of the $\as$ background.
   
\begin{figure}
\begin{center}
\includegraphics[width=75mm]{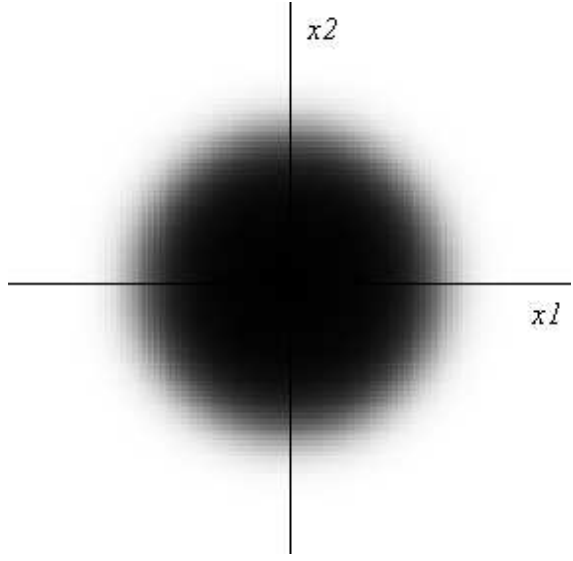}
\caption{Droplet configuration in the low temperature limit. The greyscale ring around the Fermi level corresponds to a singular region of the spacetime.}
\label{fd}
\end{center}
\end{figure}

For low temperatures, the solution is a small perturbation of the circular droplet. In fact in this limit the fermion configuration 
in the $(x_1,x_2)$  looks like a black disk with the boundary slightly ``blurred", as shown in fig. \ref{fd}.
 In the low temperature limit the expressions (\ref{z}) and (\ref{301})
 can be obtained analytically as follows \cite{Buchel:2004mc}
 \bea z^T_{FD}(R,y) & = & \frac{1}{2}+\int_0^{\infty}
n_{FD}(R')\frac{\p}{\p R'}z_0(R,y;R')dR'\cr & = &
z_0(R,y;R_0)+\frac{\pi^2 }{6} T^2 \left[\frac{\p^2}{\p
\epsilon^2}z_0(R,y;R_0=\sqrt{2\epsilon})
\right]_{\epsilon=\frac{R_0^2}{2}}
+\mathcal{O}(T^4)\label{304}\eea
and \bea V^T_{FD}(R,y) & = & -\int_0^{\infty} n_{FD}(R')g_V(R,y;R')dR'\cr
& = & V_0(R,y;R_0)-\frac{\pi^2 }{6}T^2\left[\frac{\p}{\p
\epsilon}\left(\frac{g_V(R,y;R_0=\sqrt{2\epsilon})}{\sqrt{2\epsilon}}\right)\right]_{\epsilon=\frac{R_0^2}{2}}
+\mathcal{O}(T^4)\label{305}\eea where we have used the Sommerfeld
expansion (\ref{sommerfeld}) and where \bea
V_0(R,y;R_0)=\frac{1}{2}\left(\frac{R^2+R_0^2+y^2}{\left[(R^2+R_0^2+y^2)^2-4R^2R_0^2\right]
^{1/2}}-1\right)\label{306}\eea corresponds to the $\as$ background.
In these expressions $R_0=\sqrt{2N}$ is the radius of the droplet in
the phase space at $T=0$, and $N$ is the number of fermions. It is
easy to check that (\ref{304}) and (\ref{305}) satisfy the
differential equations (\ref{diffeqpol}).

%%%%%%%%%%%%%%%%%%%%%%%%%%%%%%%

\subsubsection{ADM form of the metric}
In order to compute the mass and the angular momentum associated to
the hyperstar, it is convenient to perform the following change of
coordinates  
\bea R & = & L^2 \left(
1+\frac{r^2}{L^2}\right)^{1/2}\cos\ta\cr y & = & L \, r\, \sin\ta\cr
\phi &=& \tilde\phi + \frac{t}{L} 
\label{adscoord}
\eea 
and we also
rescale $t\rightarrow L\, t$ to have conventional units. Here $(t\,
,\, r\, ,\, \Omega_3)$ parameterize the asymptotic $AdS_5$ (in
global coordinates), whereas $(\ta\, , \, \tilde\phi\,,
\,\tilde\Omega_3)$ span the asymptotic $S^5$. Of course, $L$ is the
radius of both the $AdS_5$ and the $S^5$. It is related to the
radius $R_0$ used by \cite{Lin:2004nb} via $R_0=L^2$. In this system of
coordinates the metric can be rewritten in ADM form as 
\bea ds^2=-
\mathcal{N}^2 dt^2 + g_{\ta\ta}
\left(\frac{dr^2}{r^2+L^2}+d\ta^2\right) +g_{\tilde\phi \tilde\phi}
\left(d\tilde\phi+\mathcal{N}^{\tilde\phi}dt\right)^2
+g_{\Omega_3 \Omega_3}\ d\Omega_3^2+g_{\tilde{\Omega}_3 \tilde{\Omega}_3}\
d\tilde{\Omega}_3^2 \label{geo} 
\eea 
where $\mathcal{N}$ is the lapse function and $\mathcal{N}^{\tilde\phi}$ is the shift vector.

Introducing the expansion parameter \bea \g\equiv \frac {2 \pi^2
T^2}{3 L^8}=\frac{\pi^2 T^2}{6 N^2} \label{a2} \eea and using the
explicit expressions for $z^T_{FD}$ and $V^T_{FD}$ \bea
z^T_{FD}(R,y)&=&
\frac{R^2-R_0^2+y^2}{2\left[(R^2+R_0^2+y^2)^2-4R^2R_0^2\right]^{1/2}}+\cr&&\cr
&&+ \frac{2R_0^4
y^2\big((R_0^2+y^2)^2+R^2(R_0^2-y^2)-2R^4\big)}{\left[(R^2+R_0^2+y^2)^2-4R^2R_0^2\right]
^{5/2}}\g+\mathcal{O}(\g^2) \cr && \cr V^T_{FD}(R,y)&=&
\frac{1}{2}\left(\frac{R^2+R_0^2+y^2}{\left[(R^2+R_0^2+y^2)^2-4R^2R_0^2\right]
^{1/2}}-1\right)+\cr&&\cr &&+\frac{2R_0^4
R^2\big((R^2-R_0^2)^2-y^2(R^2+R^2_0+2y^2)\big)}{\left[(R^2+R_0^2+y^2)^2-4R^2R_0^2\right]
^{5/2}}\g+\mathcal{O}(\g^2) \label{ztfd}\eea 
one obtains, upon implementing eq. (\ref{adscoord}), the
components of the metric, which we present here up to
$\mathcal{O}(\g^2)$ terms \footnote{We notice that our expression for $g_{\tilde\phi\tilde\phi}$ differs from the one reported in \cite{Buchel:2004mc}.
} \bea \mathcal{N}^2 &=&
\left(1+\frac{r^2}{L^2}\right)\, \Big[1-\g L^2 F_1(r,\ta)\Big]
\,,\cr && \cr \mathcal{N}^{\tilde\phi} &=& \g
\frac{2L(r^2+L^2)(r^2+L^2\cos^2\ta)}{(r^2+L^2\sin^2\ta)^3} \,,\cr &&
\cr g_{\tilde\phi \tilde\phi} &=& L^2\cos^2\ta\Big[1+\g L^2
F_1(r,\ta)\Big] \,,\cr && \cr g_{\ta\ta} & = & L^2\left[1+\g
L^2\frac{r^2-L^2\sin^2\ta}{r^2+L^2\sin^2\ta} F_{2}(r,\ta)\right]\,
,\cr && \cr g_{\Omega_3 \Omega_3} & = & r^2\Big[1-\g L^2 F_2(r,\ta)\Big]
\,,\cr && \cr g_{\tilde{\Omega}_3 \tilde{\Omega}_3} & = & L^2\sin^2\ta
\Big[1+\g L^2 F_2(r,\ta)\Big] \,, \label{a1} \eea where 
\bea
F_1(r,\ta)&=&\frac{(3\cos^2\ta-1)r^4+3L^2\cos^4\ta\, r^2+
L^4(2\cos^4\ta+\sin^2\ta)}{(r^2+L^2\sin^2\ta)^3}\, , \cr && \cr
F_2(r,\ta)&=& \frac{(3\cos^2\ta-1)r^4+L^2(3 \cos^4\ta - 2\sin^2\ta) r^2+
L^4(2\cos^4\ta-\cos^2\ta-1)}{(r^2+L^2\sin^2\ta)^3}\,
\label{f1f2} 
\eea  
A general property of LLM distributions with compact support is that the 
corresponding geometries are asymptotically $\as$. One can check that this remains true for  the metric eq. (\ref{a1}). This is consistent with the fact that the droplet of fig. \ref{fd}
is effectively confined in a finite region of the phase space. From the expressions in (\ref{f1f2}), we also notice that the Sommerfeld expansion
seems no longer reliable in a region around the point $r=\ta=0$. The appearance of eventual singularities will be discussed in the following section.
%%%%%%%%%%%%%%%%%%%%%%%%%%%%%%%

\subsubsection{Singularity}

The study of singularities for LLM 
geometries was undertaken in
\cite{Caldarelli:2004mz} and \cite{Milanesi:2005} \footnote{The
resolution of singularities in this context has been investigated
in \cite{Balasubramanian:2005mg}. There, evidence was provided to
support that spacetime singularities emerge from effectively
integrating out the underlying quantum structure at the Planck
scale.}. There it was shown that all singularities appearing in the
LLM supergravity solutions are naked and fall into two classes,
namely timelike and null. While the former are considered highly
pathological due to the presence of closed timelike
curves, the latter are not. In fact, for half-BPS geometries with
null singularity, the underlying fermion density function $n(R)$
always takes values in the region $n(R)\in  (0,1)$. This is
the case both for the hyperstar and the superstar solutions.

To verify the presence of a singularity in the geometry, one should 
find a curvature invariant which diverges. The first non trivial 
invariant to consider is 
$R^2_{MN}=R_{MN}R^{MN}$, $(M,N=0,\ldots,9)$, 
since the Ricci scalar $R$ vanishes.
Indeed, as a result of the Weyl invariance
of the classical theory, the trace of the matter stress-tensor is 
identically zero. In order to check the consistency of the metric eq. (\ref{a1})
we explicitly verified that this is the case.

There are two ways to perform the computation of $R_{MN}^2$. The direct approach
involves the explicit form of the metric, while the indirect one
makes use of the field equations of type IIB supergravity. In this case, the
knowledge of the five form field strength $F_{(5)}$ will suffice:
\be\label{fieldequation} R_{MN}=\frac{1}{5!} \left(\frac{5}{2}
F_{M}^{\;\;\;P_{1}P_{2}P_{3}P_{4}}
F_{NP_{1}P_{2}P_{3}P_{4}}-\frac{1}{4}g_{MN}F^{2}\right) . \ee

Suppose now we use the approximate solution for the metric, whose
explicit form was given in the previous section, eq. (\ref{a1}). 
A lengthy calculation gives
\be \label{sing} R^{2}_{MN}=\frac{160}{L^4}+\gamma\,
\frac{P_{6}(r,\cos\theta)}{L^2 (r^2+L^2
\sin^2\theta)^4}+\mathcal{O}(\gamma^2) . \ee
In this $\gamma$ expansion, the first term 
corresponds to $AdS_{5}\times S^5$ and is, of course, finite. 
The linear term in $\g$
is, however, potentially divergent. Here $P_{6}(r,\cos\theta)$ is a
sixth order polynomial in both $r$ and $\cos\ta$ and 
goes to zero as $(r^2+L^2 \sin^2\theta)$ when $r\rightarrow 0$ 
and  $\theta\rightarrow 0$.
The square of the Ricci tensor is therefore divergent when $r=0$
and $\theta=0$. It is interesting to note here that this is
exactly the singular behavior one sees in Kerr black holes. Due to
their angular momentum, the collapsing region is not a point but a
zero-thickness ring. The Kretschmann invariant $K\sim R_{MNRS}R^{MNRS}$,
 for instance, for a
Kerr black hole with mass $M$ and angular momentum $J=M a$, is
\be K= M^2 \frac{Q_{6}(r,\cos\theta)}{(r^2+a^2 \cos^2\theta)^6} \ee
where $Q_{6}(r,\cos\theta)$ also indicates a sixth order polynomial having
the same behavior as $P_{6}(r,\cos\theta)$ in the vicinity of
$r=0$ and $\theta=0$. This is suggestive of the existence
of an event horizon in the hyperstar geometry which may manifest
itself through $\alpha'$-corrections to the supergravity solution.

This is not however the result one would have anticipated. From the
form of the metric in LLM coordinates, it is quite natural to expect
a singularity at $y=0$. Using (\ref{adscoord}), we can see that this
corresponds to $r=0$ or  $\theta=0$ in asymptotic
$AdS_{5}\times S^{5}$ coordinates. On the other hand, the singular
region appearing in (\ref{sing}) is mapped to  $(R=L,y=0)$, which is
just the Fermi surface of the fermions. We expect the singularity to be
at least smeared over an extended region around the Fermi energy, since there 
the fermion 
density is less than one, see fig. \ref{fd}.

What is therefore the true singular region of the hyperstar? We can
try to address this question in a quite general fashion valid for
all LLM geometries. We simply need to know the behavior of the
functions $z(R,y)$ and $V(R,y)$ in proximity of
$y=0$. We can distinguish two different cases depending on whether
$z_0(R,0)=\lim_{y\rightarrow 0}z(R,y)$ is independent of the radial
coordinate $R$ or not. In what follows we will focus on the latter,
since this is the case of the hyperstar.

We would like to find $z(R,y)$ and $V(R,y)$ in terms of an expansion
in $y$ or functions of $y$, such that the differential
 equations (\ref{diffeqpol}) will
be order by order satisfied. It turns out that the appropriate
Ansatz is the following
\bea z(R,y) & = & z_0(R,0) + f_{1}(R)\,y^2 \, \ln y  + \ldots \cr
V(R,y) &=& V_0(R,0) + g_{1}(R)\, \ln y  + \ldots \label{350}\eea
The functions $f_{i}(R)$ and $ g_{i}(R)$ are determined to each order from the
same differential equation (\ref{diffeqpol}). For the case that
concerns us here we have $f_{1}(R)=-\frac{1}{2 R}\p_R(R\p_R
z_0(R,0))$, $V_0(R,0)=-\frac{1}{2} R \p_R z_0$ and $g_1(R)=-R\p_R
z_0$. It is now easy to find the complete solution for
the metric and the five-form field strength in this region and
subsequently calculate $R^2_{MN}$, using either of the methods indicated
above. We find
\be R^2_{MN}=\frac{h_{1}(z_0(R,0))}{y^2}+ h_{2}(z_0(R,0),f_{1}(R))\, \ln
y +\ldots \ee
where $h_{1}(z_0(R,0))$ and $h_{2}(z_0(R,0),f_{1}(R))$ are non-zero
functions of the variables indicated.

Indeed we see that the leading term is divergent at $y=0$, as
expected. We must therefore conclude that this is the singular
region of the hyperstar, and that we cannot rely on the Sommerfeld
expansion (\ref{sommerfeld}) for calculations in the small $y$ region. This will be
important later for computing the entropy through the
Bekenstein-Hawking formula.

%%%%%%%%%%%%%%%%%%%%%%%%%%%%%%%%%%%%%%%%%%%%%%%%%%%%%%%%%%%%%%%%%%

\subsubsection{Flux}

To check the consistency of the hyperstar solution, we can verify that the flux of the 
five form $F_{(5)}$ remains equal to $N$, independently from the temperature $T$ of 
the fermion gas. From general considerations this has to be expected, since the temperature 
can be viewed as a tunable continuous parameter and as such it cannot modify the flux which 
is a topological constraint. At zero temperature, i.e. for the $\as$ solution, using the explicit expressions 
for the field strength in the LLM solution and the change of coordinates eq. (\ref{adscoord}) one 
obtains
\bea
F^{(0)}_{(5)} = \frac{r^3}{L} \, dt \wedge dr \wedge d \Omega_3 + 2 N \sin^3 \theta \cos \theta \, d \theta \wedge d \tilde{\phi} \wedge d \tilde{\Omega}_3.
\eea
The flux is computed by integrating $F^{(0)}_{(5) \theta \tilde{\phi} \tilde{\Omega}_3}$ over the $S^5$, and 
including the appropriate normalization is equal to $N$. To check that temperature perturbations do not 
alter the flux, one has to verify that corrections to $F^{(0)}_{(5) \theta \tilde{\phi} \tilde{\Omega}_3}$ 
vanish when integrated over the five sphere. Up to second order in the temperature, these take the form
\bea
F^{(1)}_{(5) \theta \tilde{\phi} \tilde{\Omega}_3} &=& \gamma \, \sin^3 \theta \cos \theta \,  
\frac{L^6 \, p_6(r,\cos \theta)}{(r^2+L^2 \sin^2 \theta)^4} \cr
F^{(2)}_{(5) \theta \tilde{\phi} \tilde{\Omega}_3} &=& \gamma^2 \, \sin^3 \theta \cos \theta \, 
\frac{L^{10} \, p_{10}(r,\cos \theta)}{(r^2+L^2 \sin^2 \theta)^8} 
\eea
where $p_6$ and $p_{10}$ are polynomials of degree 6 and 10 respectively. Although the explicit formulas look rather
involved, the integration over $\theta$ can be carried out exactly at arbitrary $r$ and indeed yields
\bea
\int_0^{\pi /2} d \theta \, F^{(1)}_{(5) \theta \tilde{\phi} \tilde{\Omega}_3} =
\int_0^{\pi /2} d \theta \, F^{(2)}_{(5) \theta \tilde{\phi} \tilde{\Omega}_3} = 0. 
\eea

%\bea F_{(2)} & = & (\p_r B_t)\, dr \wedge dt + (\p_y B_t)\, dy \wedge dt + \cr
% &+& \left[\p_r(B_t\, V)-\frac{y^3 r}{4}\p_y\left(\frac{z+\frac{1}{2}}
%{y^2}\right)\right]\, dr \wedge d\phi +
%\left[\p_y(B_t\, V)+\frac{y^3 r}{4}\p_r\left(\frac{z+\frac{1}{2}}
%{y^2}\right)\right]\, dy \wedge d\phi
%\label{311}\eea
%%%%%%%%%%%%%%%%%%%%%%%%%%%%%%%%%%%%%%%%%%%%%%%%%

\subsection{Mass}

In this section we present a systematic derivation of the ADM mass of the hyperstar solution.
The natural expectation is that this mass should coincide
with the thermal energy of the auxiliary fermion gas system.

The Einstein-Hilbert action in a $d$-dimensional spacetime is \be
S_{grav}=\frac{1}{16\pi G_{d}}\int_{\mathcal{M}}d^dx
\sqrt{-g}\,(R-2\Lambda)- \frac{1}{8\pi G_{d}}\oint_{\partial
\mathcal{M}}d^{d-1}x \sqrt{-\gamma}\,\Theta
%ds^2 = {\cal N}^2 dr^2 + \gamma_{\mu\nu}(dx^\mu +{\cal N}^\mu dr)(dx^\nu +{\cal N}^\nu dr).
\label{ADM} \ee where we included the Gibbons-Hawking boundary
term and $\gamma_{\m\n}$ is the metric on the  $(d-1)$-dimensional
timelike boundary. Following \cite{Brown:1992br}, the quasi-local
stress-tensor can be computed by the variation of the
gravitational action with respect to the boundary metric \be
T^{\mu\nu}={2\over \sqrt{-\gamma}}{\delta S_{grav}\over \delta
\gamma_{\mu\nu}}. \label{tmunu}\ee Using (\ref{ADM}) this is \be
T^{\mu\nu}={1\over 8\pi
G_d}\left(\Theta^{\mu\nu}-\Theta\gamma^{\mu\nu}
\right).\label{stress} \ee In the previous expression we have
introduced the extrinsic curvature of the $(d-1)$-dimensional timelike boundary
embedded in $\mathcal{M}$ \be \Theta^{\mu\nu}=-{1\over
2}\nabla_{(g)}^{(\mu}\hat n^{\nu)} \ee and we denoted the corresponding
 trace by $\Theta$. The covariant derivative is taken with respect to the metric $g_{\m\n}$
of the full spacetime and $\hat n^{\n}$ is the unit normal to the boundary. 
The stress-tensor (\ref{tmunu}) generically diverges as we
approach the boundary $\partial {\cal M}$ when the spacetime is
asymptotically AdS. In the context of  the AdS/CFT correspondence
we can view the gravitational quasi-local stress-tensor as the
expectation value of the stress-tensor in the associated conformal
field theory. The divergences get then a natural interpretation as
standard ultraviolet divergences  in quantum field theory
\cite{Balasubramanian:1999re}.  We can regularize the theory by
adding suitable counterterms to the original stress-tensor \be
T^{\mu\nu}={1\over 8\pi
G_d}\left(\Theta^{\mu\nu}-\Theta\gamma^{\mu\nu} +{2\over
\sqrt{-\gamma}}{\delta S_{ct}\over \delta \gamma_{\mu\nu}}\right) .
\ee
%For $AdS_5$ the right counter-terms
%\be
%T^{\mu\nu}={1\over 8\pi G}\left(\Theta^{\mu\nu}-\Theta\gamma^{\mu\nu} -{3\over l} \gamma^{\mu\nu}-{l\over 2}G^{\mu\nu}\right)
%\ee
The counterterms are consistently constructed using only  the
boundary metric $\gamma_{\m\n}$ and its covariant derivatives and are
(almost) uniquely determined by requiring a cancellation of the
divergences and general covariance (for a review, see \cite{Skenderis:2002wp}).  The boundary metric $\gamma_{\m\n}$ can be written in the
ADM form \be
\gamma_{\mu\nu}dx^{\mu}dx^{\nu}=-\mathcal{N}^2_{\Sigma}
dt^2+\sigma_{ab}(dx^a+\mathcal{N}^a_{\Sigma}
dt)(dx^b+\mathcal{N}^b_{\Sigma} dt) \ee where $\Sigma$ is a
surface of constant $t$ inside $\partial {\cal M}$. Conserved
charges are obtained by integrating $T^{\mu\nu}$ over a spacelike
hypersurface at infinity. A finite expression for the mass is
obtained substituting the regularized stress-energy tensor in the
following formula \be
M=\int_{\Sigma}d^{d-2}x \, \sqrt{\sigma}\mathcal{N}_{\Sigma}u^{\mu}T_{\mu\nu}u^{\nu} ,
\ee where $u^{\mu}$ is the timelike unit normal to $\Sigma$. For instance, the application of this method to the
five-dimensional AdS-Schwarzschild black hole \be ds^2=-\left[{r^2
\over L^2} +1-\left({r_0 \over r}\right)^2 \right]dt^2 + {dr^2
\over \left[{r^2 \over L^2} +1-\left({r_0 \over r}\right)^2
\right]} +r^2(d\theta^2 + \sin^2{\theta}d\phi^2 + \cos^2{\theta}
d\psi^2) \ee
 yields \cite{Balasubramanian:1999re}
\be M={3\pi l^2\over 32 G_5}+{3\pi r_0^2\over 8 G_5}. \ee The first
term,  which is present also when the black hole disappears,
corresponds to the Casimir energy of the vacuum in the dual CFT.

It would be nice to have a similar counterterm  method directly in
a ten-dimensional setting. Unfortunately, extending
the program of holographic renormalization to ten-dimensional
metrics with $AdS_5\times S^5$  asymptotics  seems problematic
\cite{Taylor-Robinson:2001fe}. We are therefore forced to use
alternative approaches. In the first one, we will determine the
relevant components of the stress-tensor relative to some
reference geometry following \cite{Myers:1999ps}. The second
approach is the so called background subtraction method
\cite{Hawking:1995fd}. In both cases one has to carefully match
the asymptotic geometry of the supergravity solution with that of
a reference background. Neither of the methods  can reproduce
 the Casimir energy of the associated CFT. However this
will not be a problem in our case since we are interested in
computing the energy difference between the half-BPS supergravity
solution and the $AdS_5\times S^5$ ground state.

We now proceed to compute the mass of the hyperstar (\ref{a1}) as a
series expansion in the small parameter $\gamma\equiv \frac{\pi^2 T^2}{6 N^2}$. This mass should agree with the energy of the free
fermion gas, eq. (\ref{smallTen}). We will 
first consider the leading order in $\g$ and comment on $\g^2$ orders in a later section.

%%%%%%%%%%%%%%%%%%%%%%%
%%%%%%%%%%%%%%%%%%%%%%%%%%%%%%%%%%%%%%%%%%%%%%%

\subsubsection{First approach}

Following \cite{Myers:1999ps}, we 
obtain the stress-tensor associated with the metric (\ref{a1}) relative to the
$\as$ background metric $g^0_{\m\n}$. 
We need to require that the difference
between the two metrics falls off suitably fast for large radius.
Explicitly we want that \bea
g_{rr}-g_{rr}^0 = o(1/r^6)\cr g_{ra}-g_{ra}^0 =
o(1/r^5)\label{321}\eea where $o(1/r^n)$ means that these
differences go to zero more rapidly than $1/r^n$ and the index $a$
runs over all the coordinates except $r$.
To satisfy such requirement we implement an appropriate
change of coordinates $(r,\ta)\rightarrow (\tilde r, \tilde\ta)$, which we presently discuss.
The effect of using these new
coordinates is to make the leading asymptotic perturbations of the
metric all in components parallel to the boundary directions. Then
the line element becomes \bea ds^2 = g_{\m\n}^0
dx^{\m}dx^{\n}+\frac{\hat{T}_{ab}}{\tilde r^2}dx^{a}dx^{b}+\ldots\label{322}\eea
from which one can read off the stress-tensor up to a multiplicative
constant depending only on the space-time dimensions.

The first step is therefore to find a coordinate system such that
the metric satisfies (\ref{321}). We consider the Ansatz 
\bea  r  &=& \tilde{r}+\frac{f_1}{\tilde{r}}+\frac{f_2}{\tilde{r}^3} \cr
\cos{\theta} & = &
\cos{\tilde{\theta}}+\frac{f_3}{\tilde{r}^2}+\frac{f_4}{\tilde{r}^4}\label{323}
\eea
where the  $f_i=f_i(\tilde\ta)$ $(i=1,\ldots, 4)$ are functions to be determined in 
order to  adjust the asymptotics of the metric. 
%The powers of $\tilde{r}$ go in steps of 2....

In terms of the new variables $\tilde r$ and $\tilde\theta$, the $g_{\tilde r \tilde r}$
component of the metric has an expansion for large $\tilde{r}$ which
differs from the background reference metric
$g^0_{\tilde r \tilde r}=(1+\tilde{r}/L^2)^{-1}$ by terms containing the $f_i$.
The $1/\tilde{r}^4$ term can be eliminated by tuning
$f_1=\frac{1}{4}(3 \cos{\tilde{\theta}}^2-1)L^2 \g$ and similarly the $1/\tilde{r}^6$ 
 with an
appropriate choice of $f_2$. The first constraint in
(\ref{321}) is then satisfied. Analogously, $f_3$ and $f_4$ are fixed by
requiring the vanishing of the $1/\tilde{r}^3$ and $1/\tilde{r}^5$
terms in $g_{\tilde r \tilde\theta}$, which appears after
changing variables according to (\ref{323}). Once the $f_i$ are fixed, one can verify
that the other components of the metric coincide with
$g^0_{\m\n}$ up to orders $\mathcal{O}\left(\tilde{r}^{-2}\right)$.
The only disagreement
is found in $g_{tt}$ and $g_{\Omega_3\Omega_3}$, which contain
a term at order $\mathcal{O}\left(\tilde{r}^0\right)$
\bea 
\g(3\cos^2\tilde\ta-1) 
\eea
which, nonetheless, vanishes upon integration over the $S^5$ (including the appropriate measure). We notice that the 
same factor already appeared at leading order in the asymptotic expansion of the metric perturbation, see eq. (\ref{f1f2}).

From (\ref{322}) and the explicit expression for \bea
g_{tt}=-\mathcal{N}^2+g_{\tilde\phi\tilde\phi} \left(
\mathcal{N}^{\tilde\phi} \right)^2 \label{gtt}\eea one can read off
the time-time component of the stress-tensor \bea \hat{T}_{tt} & = &
\left(g_{tt}(\tilde{r},
\tilde{\theta})+1+\frac{\tilde{r}^2}{L^2}\right)\tilde{r}^2\cr &&\cr
& = & \frac{\g}{8}\left(4(3\cos^2\tilde\ta-1)\tilde{r}^2+L^2\big(11
- 39 \cos^2\tilde\ta+60\cos^4\tilde\ta\big)\right)+
\mathcal{O}\left(\frac{1}{\tilde{r}^2}\right) . \label{324}\eea This
expression has to be integrated at the spacelike boundary
in order to give the mass \bea M &=& \frac{4}{16 \pi G_{10}}
\int \, \hat{\mu} \, \hat{T}_{tt}\cr  &=& \frac{4}{16 \pi G_{10}} L^5 (2
\pi) (2 \pi^2)^2 \int _0^{\pi/2} d\theta \cos{\theta} \sin^3{\theta}
\; \hat{T}_{tt}\label{325}\eea where $G_{10}=
\frac{\pi^4 L^8}{2 N^2}$ and $\hat{\mu}=\tilde{r}^{-3} \sqrt{g_{\tilde\phi\tilde\phi}\,
g_{\ta\ta}\, g_{\Omega\Omega}^3\,
g_{\tilde{\Omega}\tilde{\Omega}}^3} $ is the integration measure. The
final result for the mass is \bea M = \frac{L^7 }{4}\g  =
\frac{\pi^2}{6 L}T^2\label{326}\eea which agrees with the thermal
excitation energy of the $N$ fermions above the ground state, eq.
(\ref{smallTen}). The extra $L$ in the denominator comes from the
rescaling of the time variable already discussed.
It is important to remark that in obtaining these expressions we have consistently 
worked at order $\g$. We will comment on the significance
of higher order terms in a later section.

%%%%%%%%%%%%%%%%%%%%%%%%%%%%%%%%%%%%%%%%%%%%%%%

\subsubsection{The superstar}

As a further check of the validity of the procedure just discussed,
we also apply it to the so-called \emph{superstar}, a family of
asymptotically $\as$ backgrounds discovered in \cite{Myers:2001aq}
and further studied from the LLM perspective in \cite{Suryanarayana:2004ig} \cite{Caldarelli:2004mz}
\cite{Shepard:2005zc} \cite{Balasubramanian:2005mg}.
 The extremal 1/2 BPS superstar metric is
governed by two parameters, the flux $N$ of the 5-form through the
$S^5$ and one of the three angular momenta on the $S^5$, $J_3$, which
coincides with the energy $\Delta$ because of the BPS condition.
Explicitly the metric can be written as \cite{Caldarelli:2004mz} \bea ds^2 &=&
-\frac{1}{G}\left(\cos^2\ta+\frac{r^2}{L^2}G^2\right)dt^2 +
\frac{L^2 H}{G} \sin^2\ta d\phi^2 + 2\frac{L}{G}\sin^2\ta dt d\phi
+\cr && \cr &&+ G\left(\frac{dr^2}{f}+r^2 d\Omega_3^2\right)+L^2G
d\ta^2 +\frac{L^2}{G}\cos^2\ta d\tilde\Omega_3^2
\label{superstar}\eea with \bea f = 1+H\frac{r^2}{L^2}\, , ~~~~~~ G
 = \sqrt{\sin^2\ta + H \cos^2\ta}\, , ~~~~~ H &=&
1+\frac{2L^2\Delta}{N^2 r^2}\equiv
1+\frac{Q}{r^2} . \label{superstar2}
\eea 
Also in this example we want
to satisfy  the fall off conditions (\ref{321}). By choosing an
appropriate coordinate system as in (\ref{323}) it is easy to see
that \bea \hat{T}_{tt}& = &
\frac{Q}{4L^2}\big(2-3\cos^2\tilde\ta\big)^2\tilde{r}^2+ \cr &&\cr
&&+\frac{Q}{64
L^2}\big((6-36\cos^2\tilde\ta)L^2-(4+15\cos^2\tilde\ta-24\cos^4\tilde\ta)Q\big)
+\mathcal{O}\left(\frac{1}{\tilde r}\right) .
\label{Thatttsuperstar}\eea The expression for the mass is then \bea
M &=& \frac{4}{16 \pi G_{10}} \int \, \hat{\mu} \, \hat{T}_{tt}=
\frac{4}{16 \pi G_{10}} L^5 \pi^5 Q =
\frac{\Delta}{L}\label{masssuperstar}\eea which, up to the $L$
coming from the rescaling of the time, is exactly the energy of the
geometry. Note that we have again neglected contributions quadratic in $Q$ 
in the stress-tensor (\ref{Thatttsuperstar}).

%%%%%%%%%%%%%%%%%%%%%%%%%%%%%%%%%%%%%%%%%%%%%%%%%%%%%%%%%

%%%%%%%%%%%%%%%%%%%%%%%%%%%%%%%%%%%%%%%%%%%%%%%%%%%%%%%%%

\subsubsection{Second approach: Background subtraction}

We now discuss the second approach \cite{Hawking:1995fd} for computing the mass of the
hyperstar. In the background subtraction prescription the ADM mass
is obtained by integrating the quasi-local energy ${\cal
N}(K-K_0)$ over the $(d-2)$-dimensional spacelike 
hypersurface $\Sigma$ at radial infinity \bea
M=\frac{1}{8\pi G_{10}}\int_{\Sigma}\,\mu\, \mathcal{N}\,(K-K_0).
\label{masshh1}\eea To obtain $M$ one needs $K^{\m\n}$, the extrinsic
curvature of $\Sigma$ embedded in a constant time hypersurface 
\bea K^{\mu\nu}=-{1\over
2}\nabla_{(h)}^{(\mu}\hat r^{\nu)} .
\eea  Now the covariant derivative is calculated with respect to the metric $h_{\mu\nu}$ of the constant time hypersurface, and 
$\hat r^{\n}=g_{rr}^{-1/2}\delta^{\n}_r$. In (\ref{masshh1})
$K$ and $K_0$ are the traces of the extrinsic curvature of the 
spacetime and of the
reference background respectively, and $\mu$ is the measure on $\Sigma$.

In this case we also need to carefully tune the components of the
boundary metric with those of the $\as$ background by performing
an asymptotic coordinate transformation as in the Ansatz
(\ref{323}). Let us consider the extremal superstar solution in
its five-dimensional reduction  to understand which fall-off
requirements we need to impose. The mass of this solution was first 
obtained in \cite{Behrndt:1998jd}. The line element reads \be
ds^2=-H^{2/3}f dt^2+H^{1/3}\left(f^{-1} dr^2+r^2
d\Omega_3^2\right)\ee where 
%\be H=1+{Q\over r^2},\,\,\,\,f=1+{r^2\over L^2}H. \ee 
$H$ and $f$ are defined as in (\ref{superstar2}).
The parameter $Q$ appearing in (\ref{superstar2}) is the
five-dimensional electric charge and corresponds to the angular
momentum $J$ in the ten-dimensional uplifting of the superstar
solution (\ref{superstar}), see also \cite{Cvetic:1999xp}. We perform the following change of
variable on the solution \be {\tilde r}^2={ r}^2H^{1/3} \ee which
asymptotically amounts to \be r= \tilde r -{Q\over 6\tilde
r}.\label{asi} \ee 
A posteriori one can verify that additional higher order terms in (\ref{asi})
 do not modify the final answer for the mass. 
After this
transformation, the difference between the components of the
boundary metric $g_{\Omega_3\Omega_3}$
 and the global
$AdS_5$ background becomes of order ${\cal O}(\tilde r^{-4})$. An
explicit calculation of the extrinsic curvature yields \be
K=-{3\over L}-{3L\over 2\tilde r^2}+ \left({3 L^3\over
8}-{Q^2\over 3L}+LQ\right){1\over \tilde r^4}+{\cal
O}\left({1\over \tilde r^6}\right).  \label{Ksup} \ee To obtain a finite mass,
we need to subtract the extrinsic curvature of $AdS_5$ \be
K_0=-{3\over L}\left(1+{L^2\over \tilde r^2}\right)^{1/2}. \ee
Using \bea M=\frac{1}{8\pi G_{5}}\int\,\mu\, \mathcal{N}\,(K-K_0)
\label{masshh2}\eea with $G_5=\pi/4$ \footnote{In this example we 
use the units of \cite{Behrndt:1998jd}.}, we obtain the well known
result \be M={Q }.\label{BPS} \ee One can easily verify that if we
had not implemented the transformation (\ref{asi}) we would have
gotten \be K=-{3\over L}-{3L\over 2\tilde r^2}+ \left({3 L^3\over
8}-{Q^2\over 3L}+{3LQ \over 2}\right){1\over \tilde r^4}+{\cal
O}\left({1\over \tilde r^6}\right) \ee and correspondingly the
incorrect result \be M={3\over 2}\,Q. \ee 
As in the ten-dimensional example (\ref{superstar}), we have again
neglected a term
proportional to $Q^2$.

 We now proceed similarly with the hyperstar solution using $f_1$, $f_2$ to fix
 the asymptotic behavior
of $g_{\Omega_3 \Omega_3}$ and analogously $f_3$, $f_4$ to fix
$g_{\tilde{\Omega}_3 \tilde{\Omega}_3}$ 
\footnote{We could have also chosen to use the parameters $f_i$ to fix the other components $g_{\phi\phi}$ ,$g_{\theta\theta}$ of the boundary metric. This ambiguity  alters only the quadratic contribution
to the mass which, as will be discussed, is not physical.}. Having four parameters at
our disposal we require that
 $\delta g_{\Omega_3 \Omega_3}$ and $\delta g_{\tilde\Omega_3 \tilde\Omega_3}$ are of order
  ${\cal O}\left({\tilde
r^{-4}}\right)$.  With this choice we obtain $\delta
g_{\ta\ta}={\cal O}(\tilde r^{-2})$, while for the other
component of the boundary metric $g_{\tilde\phi\tilde\phi}$ we have $\delta
g_{\tilde\phi\tilde\phi}=\gamma L^4(-1+3\cos(\tilde \ta))$ which integrates to
zero on the $S^5$.
  %We can  \be f_1={1\over
%2}(3\cos(\tilde\ta)-1)L^2\gamma \ee
After having implemented this coordinate transformation, we can
compute the extrinsic curvature to linear order in $\g$ 
obtaining \bea K &=&
-\frac{3}{L}-\frac{L}{2\tilde{r}^2}\Big(3-7\big(3\cos^2\tilde\ta-1\big)\g\Big)+\cr&&
\cr &&+
\frac{L^3}{8\tilde{r}^4}\Big(3+4\big(28-159\cos^2\tilde\ta+174\cos^4\tilde\ta\big)\g
\Big) +
\mathcal{O}\left(\frac{1}{\tilde{r}^5}\right) \label{k}.\eea  
Subtracting the extrinsic curvature
contribution of the background \bea
K_0=-\frac{3}{L}\left(1+\frac{L^2}{\tilde{r}^2}\right)^{1/2} =
-\frac{3}{L}-\frac{3L}{2\tilde{r}^2}+\frac{3L^3}{8\tilde{r}^4}+\mathcal{O}\left(\frac{1}{\tilde{r}^5}\right)
\label{k0}\eea
%and with the measure \bea
%\mu=\Big(g_{\tilde\phi\tilde\phi}\, g_{\ta\ta}\,
%g_{\Omega\Omega}^3\,
%g_{\tilde{\Omega}\tilde{\Omega}}^3\Big)^{1/2}\label{measure}\eea
and using the ADM mass formula, eq. (\ref{masshh1}), we obtain
\bea M = \frac{L^7}{4}\g = \frac{\pi^2 }{6 L}T^2\label{336}\eea
which is again the expected result.

%%%%%%%%%%%%%%%%%%%%%%%%%%%%%%%%%%%%%%%%%%%%%%%%%%%%%%%%%%%

\subsubsection{Contributions to the mass of order $\g^2$}

It remains to discuss the relevance of the quadratic terms
in $\g$ that we have so far consistently neglected.
According to the discussion following eq. (\ref{smallTen}), we would
not expect contributions to the mass at orders higher than $\g \sim
T^2$. We now check whether this is the case. Using the expressions at
order $\g^2\sim T^4$ for $z^T_{FD}$ and $V^T_{FD}$ \bea
z^{T\;\;(2)}_{FD}(R,y)&=&\g^2\frac{84 R_0^8\, y^2}
{5\left[(R^2+R_0^2+y^2)^2-4R^2R_0^2\right]^{9/2}}\cdot \cr&& \cr
&&\cdot \Big((R_0^2+y^2)^4+R^2(R_0^2-11y^2)(R_0^2+y^2)^2 + \cr
&&\hskip 2.5cm + 3
R^4(3R_0^4+3R_0^2y^2-2y^4)+R^6(11R_0^2+14y^2)-4R^8\Big)\cr && \cr
V^{T\;\;(2)}_{FD}(R,y)&=& \g^2\frac{84 R_0^8\, R^2}
{5\left[(R^2+R_0^2+y^2)^2-4R^2R_0^2\right]^{9/2}}\cdot \cr&& \cr
&&\cdot \Big((R_0^2-4y^2)(R_0^2+y^2)^3-R^2(4
R_0^6+9R_0^4y^2-9R_0^2y^4-14y^6) +\cr  && \hskip 2.5cm +
R^4(6R_0^2+21R_0^2y^2+6y^4)-R^6(4R_0^2+11y^2)+R^8\Big)\label{zsecond}\eea
it is straightforward to write down the corresponding asymptotic expression
for large $r$ of the hyperstar metric, which is not particularly illuminating and, therefore, we do not present it.   

It is not difficult to see that, differently from what expected, there seems
to be a non vanishing contribution to the mass proportional to
$\g^2$ \footnote{On the other hand, the angular momentum, which can be obtained from ${\cal N}^{\tilde\phi} \sim {\g \over \tilde{r}^2} + {\g^2 \over \tilde{r}^4} + \ldots$, does not receive corrections beyond ${\cal O} (\g)$.}. The exact coefficient of this term depends on the procedure
used to compute it. In the first approach discussed above there is 
a quadratic contribution to the stress-tensor 
\bea \hat{T}_{tt}^{(2)}=-\frac{\g^2}{16}\big(19-159\cos^2\tilde\ta+216
\cos^4\tilde\ta\big)\eea  and to the mass \bea
M^{(2)} = - \frac{L^7}{32}\g^2 = -\frac{\pi^4 }{72 L^9}T^4 . 
\label{m2myers}\eea The method of background subtraction
gives \bea K^{(2)}= -\g^2\frac{L^3}{8\tilde{r}^4}
\big(69-540\cos^2\tilde\ta+747\cos^4\tilde\ta\big)\eea 
and \bea M^{(2)} = - \frac{125 L^7}{128}\g^2 = -\frac{125 \pi^4
}{288 L^9}T^4 . \label{m2hh}\eea

The presence of this term and its scheme dependence are, however,
not completely surprising and have already been discussed in the
literature. In computing the superstar mass, 
both in five and ten dimensions, we already encountered a similar issue, see eqs. (\ref{Thatttsuperstar}) and (\ref{Ksup}). Indeed,
retaining the $Q^2$ terms in the computation of the mass, one would   
obtain a non linear BPS condition $M \simeq Q
-\frac{Q^2}{3L^2}$
\cite{Buchel:2003re}. This
relation clearly conflicts with the expectation $M\geq |Q|$. One can
nevertheless recover the usual linear BPS condition by including
appropriate finite counterterms related to scalar fields
\cite{Liu:2004it}.
This discussion can be generalized to the three-charged $AdS_5$ black hole.
It has been observed in \cite{Caldarelli:2004ig} that terms
quadratic in the charges are related to a trace
anomaly of the stress-tensor. 
%\bea 8\pi G_5
%T_{\m}^{\;\;\m}=-\frac{1}{L^3
%r^2}\left(\sum_{i<j}Q_iQ_j-\frac{1}{3}\left(\sum_i
%Q_i\right)^2\right)g_{\m}^{\;\;\m}+\mathcal{O}\left(\frac{1}{r^4}\right)\eea
This anomaly stems from a renormalization scheme which violates the
asymptotic isometry group of $AdS_5$ and can be removed by adding to
the action the finite counterterm proposed in \cite{Liu:2004it}.

In the light of these examples, we therefore consider
(\ref{m2myers}) and (\ref{m2hh}) as spurious: They should be eliminated 
by a convenient choice of counterterms, although we do not know how to carry out 
this procedure directly in ten dimensions. 

Orders beyond $\g^2$ do not contribute to the mass of the solution, because they fall off 
too fast at radial infinity.

%%%%%%%%%%%%%%%%%%%%%%%%%%%%%%%%%%%%%%%%%%%%%%%%%%%%%%%%%%%%
\subsection{Angular momentum}

As a check of the BPS condition for the hyperstar solution, we now calculate  the associated angular momentum.
This computation is most easily done in a five-dimensional setting.  The ten-dimensional angular momentum $J$ 
coincides with the electric charge $Q$ of the $U(1)$ gauge field $A$ coming from dimensional reduction on the $S^5$. 
The gauge field can be read off from the term $g_{\tilde\phi \tilde\phi}(d\tilde\phi+{\cal N}^{\tilde\phi}dt)^2$  
in the ADM metric and therefore coincides with the shift vector:
\be 
A={\cal N}^{\tilde\phi}\, dt={2L\over r^2}\gamma \, dt +{\cal O}\left({1\over r^4}\right).
\ee
The associated charge (angular momentum) is then 
\footnote{Looking at the five-dimensional gauged supergravity action one 
would expect a contribution  to the charge of the type $ \int_{S^3_\infty} A\wedge F $. 
This term is nonetheless subleading and vanishes at radial infinity. }
\be
J={L^2\over 16\pi G_5}\int _{S^3_\infty}\star_5\, dA=\gamma {L^8\over 4}= \frac{\pi^2}{6}T^2
\label{angular}
\ee
where $\star_5$ is the five-dimensional Hodge star operator. 
In our normalization the five-dimensional Newton constant is $G_5=G_{10}/Vol(S^5)=2\pi/L^5.$ The $L^2$ factor 
in (\ref{angular}) is necessary for obtaining conventional units. Comparing $J$ with the mass formula eq.
(\ref{326}), we obtain the BPS relation $M=J/L$.

%%%%%%%%%%%%%%%%%%%%%%%%%%%%%%%%%%%%%%%%%%%%%%%%%%%%%%%%

\subsection{Entropy}

%%%%%%%%%%%%%%%%%%%%%%%%%%%%%%
%Expansion for small $y$
%We assume the following Ansatz for the expressions of $z$ and $V$ at small $y$
%\bea z(R,y) & = & z_0(R,0) + f(R)\,y^2 \, \ln y  + \ldots \cr
%V(R,y) &=& V_0(R,0) - g(R)\, \ln y  + \ldots \label{350}\eea where
%the functions $V_0$, $f$, and $g$ are to be determined from the differential equations (\ref{diffeqpol}).
%It is easy to see that $V_0=-\frac{1}{2} R \p_R z_0$, $f(R)=-\frac{1}{2 R}\p_R(R\p_R z_0)$,
%and $g(R)=R\p_R z_0$.
%%%%%%%%%%%%%%%%%%%%%%%%%%%%%%

In the previous sections we have found agreement between the ADM mass and the thermal energy of the fermions.
Since the Fermi gas has non-vanishing entropy at non-zero temperature, we expect the same to occur for the supergravity solution. We would like to understand how this entropy arises geometrically in the case of the hyperstar. 
Although the solution we are considering seems to have a naked
singularity, it is expected that $\alpha '$ corrections to the
equations of motion might generate a finite-area \emph{stretched horizon}. With these corrections we can think of the hyperstar as a
legitimate black hole.

In the presence of an event horizon, the entropy of a gravitational
solution in $d$ dimensions is given by the celebrated
Bekenstein-Hawking formula \bea S = \frac{\mathcal{A}_d}{4
G_{d}}\label{351}\eea where $\mathcal{A}_d$ is the area of the
horizon. In our case the entropy is still given by
(\ref{351}) but now $\mathcal{A}=\mathcal{A}_{sh}$ is the area of
the stretched horizon.

Since we do not know the explicit form of the $\alpha'$ corrections, the location of the stretched horizon is inherently ambiguous. Therefore we expect to reproduce the fermion entropy up to a numerical coefficient.

 As we already discussed, the $y=0$ plane is a null singular region. It is reasonable to assume
that the $\alpha'$ corrections will generate a horizon at $y_{sh}\simeq 0
+\mathcal{O}(\alpha ')$.   

We therefore need to compute the area of the $y=y_{sh}$ plane, with $y_{sh} \simeq \alpha '
= g_s^{-1/2} \, l_p^2 \sim g_s^{-1/2}$ in units where $\hbar = 1$. This area
turns out to be finite. The metric in LLM coordinates for fixed $t$ and $y$ reads \bea
ds^2|_{t\, ,\,y\,=\,fixed}=-h^{-2}V^2d\phi^2 +
h^2(dR^2+R^2d\phi^2)+y e^G d\Omega_{3}^2+y e^{-G}
d\tilde{\Omega}_3^2\label{352}\eea so that the integration measure is
\bea \m=\sqrt{h^4 R^2-V^2}\, y^3\simeq h^2\,  R \,
y^3\simeq\left(\frac{1}{4}-z_0^2\right)^{\frac{1}{2}} \, R\,
y^2\label{353}\eea where we have assumed the expansion (\ref{350}), so
that the term $V^2 \sim \ln^2 y$ can be neglected for small $y$ against $h^4
\sim y^{-2}$ and  $z \simeq z_0$. By restricting the measure (\ref{353}) 
to $y=y_{sh}$, the Bekenstein-Hawking formula yields 
\bea S & = & \frac{\mathcal{A}_{sh}(y=y_{sh})}{4 G_{10}}=\frac{2\pi
(2 \pi^2)^2}{4 \cdot 2\pi^4}y_{sh}^2 \int_0^{\infty} R dR
\left(\frac{1}{4}-z_0^2\right)^{1/2}\cr & \simeq & c
\int_0^{\infty} R dR \sqrt{n(1-n)} \label{354}
\eea
where $c$ is a numerical constant. For the hyperstar $n=n_{FD}$ so that 
\bea 
\sqrt{n_{FD}(1-n_{FD})}=
\frac{e^{\frac{\beta}{2}(\epsilon - \m)}}{1+
e^{\beta(\epsilon-\m)}}\label{355}
\eea 
with $\epsilon=R^2/2$ and $\m=T \left(e^{N/T}-1\right)$. Using eq. (\ref{354}) we obtain  
\bea 
S  & \simeq & c \int_0^{\infty}
d\epsilon \frac{e^{\frac{\beta}{2}(\epsilon - \m)}}{1+
e^{\beta(\epsilon-\m)}} \cr & = & 2 \,c \, T \left(
\frac{\pi}{2}-\arctan e^{-\frac{\beta\m}{2}}\right) = 2 \, c \, T
\left( \frac{\pi}{2}-\arctan
\frac{1}{\sqrt{e^{N/T}-1}}\right) .
\label{356}
\eea 
In the low temperature approximation this yield 
\bea S \propto T\left(1+
\mathcal{O}(e^{-N/T})\right).
\label{357}
\eea
Therefore the entropy is proportional to $T$, as expected from eq. (\ref{entrosmall}), up to corrections which
are exponentially suppressed for $N\gg T$. 

In the high temperature limit, however, eq. (\ref{356}) does not seem to reproduce the logarithmic behavior 
of the Boltzmann entropy. In this limit, which will be studied in the next section, the assumptions and the approximations 
which led to eq. (\ref{354}) might not be valid since $T$ is not a small parameter.

%%%%%%%%%%%%%%%%%%%%%%%%%%%%%%%%%%%%%%
\section{High temperature regime}
%%%%%%%%%%%%%%%%%%%%%%%%%%%%%%%%%%%%%

\begin{figure}

\begin{center}

\includegraphics[width=75mm]{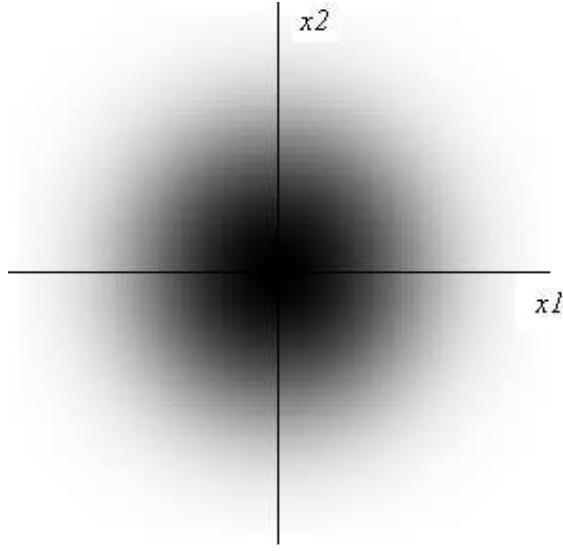}

\caption{Droplet configuration in the high temperature limit.}
\label{bol}

\end{center}

\end{figure}

We now move to consider the high temperature regime. In this limit
the Fermi-Dirac distribution reduces to the classical Boltzmann
distribution. Correspondingly, the droplet spreads over a larger part of the $y=0$ 
plane and the singular greyscale region is not confined inside a thin ring anymore, as shown in fig. \ref{bol} .   

The auxiliary function $z$ can be computed in this regime as 
\bea 
z^T_{B}(R,y) & = & \frac{1}{2}+\int_0^{\infty}
n_{B}(R')\frac{\p}{\p R'}z_0(R,y;R')dR'
\eea
where 
\bea
n_{B}(R)= N\beta e^{-\beta {R^2 \over 2}} 
\eea 
and $\beta = 1/T$.
Making the change of variable $R'^2 /2 = \epsilon$ and using the explicit expression for $z_0(R,y;R')$ 
we can write the integral as
\bea 
z^T_{B}(R,y) = {1\over
2}-2y^2 N \beta \,I_z(\beta,R,y) 
\label{zTB}
\eea 
where we have defined 
\bea
I_z(\beta,R,y) \equiv 
 \int_0^{\infty}d\epsilon \; e^{-\beta
\epsilon}{(2\epsilon+y^2+R^2)\over
[(2\epsilon+y^2+R^2)^2-8R^2\epsilon]^{3/2}}. 
\label{Iz}
\eea
The high temperature limit corresponds to the small $\beta$ region. Therefore we 
want to find an approximate expression for $I_z(\beta,R,y)$ near $\beta = 0$. 
It is easy to
verify that $I_z(0,R,y)={1 \over 2y^2}$, and also that 
\bea 
{\partial I_z\over \partial \beta} = -\int_0^{\infty}d\epsilon
\;{\epsilon \; e^{-\beta
\epsilon} (2\epsilon+R^2+y^2)\over
[(2\epsilon+y^2+R^2)^2-8R^2\epsilon]^{3/2}} 
\eea 
diverges as $\ln\beta$ in proximity of $\beta = 0$, because the integrand goes like $e^{-\beta
\epsilon} / \epsilon$ for large $\epsilon$. This suggests a low
$\beta$ expansion of the form 
\bea
I_z(\beta,R,y)&=&I_z(0,R,y)+A\beta\ln\beta+...\nonumber\\
z^T_B(R,y)&=&{1\over 2}-2y^2N\beta \left({1\over
2y^2}+A\beta\ln\beta+...\right) . 
\eea 
Since it is not possible to compute explicitly ${\partial I_z\over
\partial \beta}$ for $\beta \rightarrow 0$ because of the divergence, 
to find its small $\beta$ behavior we find it useful to 
first regulate the integral  by considering the quantity 
\bea
{\partial I_z\over
\partial \beta}+{1\over 4}\int_0^{\infty}d\epsilon \;{\epsilon\,
e^{-\beta\epsilon}\over\epsilon^2+(R^2+y^2)^2}. \label{combi}
\eea
The new piece 
\bea 
I_0 \equiv -{1\over 4}\int_0^{\infty}d\epsilon
\;{\epsilon\, e^{-\beta\epsilon}\over\epsilon^2+(R^2+y^2)^2} 
\eea
has the same divergence structure of ${\partial I_z\over \partial
\beta}$ and its value is known for finite $\beta$ in terms of the Sine and Cosine Integral functions 
$Si(x)$, $Ci(x)$.  The
corresponding small $\beta$ expansion can be given explicitely as 
\bea 
I_0 = \, {\gamma\over
4}+{1\over 4}\ln[\beta(R^2+y^2)]-{\pi\over 8}\beta(R^2+y^2)+{\cal
O}(\beta^2 \ln \beta)\label{Izero} 
\eea
where $\gamma$ is the Euler-Mascheroni constant.
The combination (\ref{combi}) is by construction convergent for any $\beta$,
and has a well defined $\beta \rightarrow 0$ limit which can be easily computed analytically 
\bea 
\left({\partial I_z\over
\partial \beta}-I_0\right) \Big{|}_{\beta = 0}={1\over 4}\left(1-{R^2\over y^2}\right)-{\ln 2\over 4}
+{1\over 4}\ln y^2-{1\over 4}\ln(R^2+y^2). 
\eea 
Using eq. (\ref{Izero}) we obtain the high temperature expansion of
${\partial I_z\over \partial \beta}$ 
\bea
{\partial I_z\over \partial \beta} = {1 \over 4} \ln \beta + {1 \over 4}\left(\gamma -\ln 2 + 1-{R^2\over
y^2}  + \ln y^2 \right) + {\cal O}(\beta \ln \beta)
\eea
and integrating in $\beta$ we can finally get 
\bea 
I_z(\beta,R,y) = {1 \over 2y^2} + {1\over 4}\beta \ln\beta+{\beta\over 4}\left(\gamma-\ln 2-{R^2\over
y^2}  + \ln y^2 \right) + {\cal O} (\beta^2 \ln \beta).
\eea 
The corresponding high temperature limit of $z^T_B$, keeping only the first two orders, reads then as follows 
\bea 
z^T_B (R,y) = {1\over 2}-N\beta-{ N y^2\over
2}\beta^2\ln\beta +  {\cal O}(\beta^2) .
\label{zhigh}
\eea

To obtain the metric we need to find also the function $V^T(R,y)$. Starting 
from eq. (\ref{301}) and inserting the Boltzmann distribution we arrive at 
\bea 
V^T_B(R,y)= \int_0^{\infty}d\epsilon \; N\beta e^{-\beta
\epsilon}{2 R^2(2\epsilon+y^2+R^2)\over
[(2\epsilon+y^2+R^2)^2-8R^2\epsilon]^{3/2}} \equiv 2N \beta R^2 I_V(\beta,R,y). 
\label{VTB}
\eea 
One can verify that $I_V(0,R,y)$ vanishes and that ${\p
I_V\over\p \beta}$ diverges logarithmically in the $\beta \rightarrow 0$ limit. We can proceed
similarly as before by regulating ${\p
I_V\over\p \beta}$ with an appropriate ``reference" integral, to finally obtain 
\bea 
{I_V(\beta,R,y)} = -{1 \over 4} \beta \ln \beta -{\beta \over 4} \left(\gamma -\ln 2 + 1 + \ln y^2 \right) + {\cal O}(\beta^2 \ln \beta).
\eea
In proximity of $\beta=0$ the leading
contribution to $V^T_B$ is therefore 
\bea 
V^T_B(R,y)=-{1\over
2}N R^2 \beta^2 \ln\beta + {\cal O}(\beta^2).
\label{Vhigh} 
\eea 
The expressions for $z^T_B$ and $V^T_B$ consistently satisfy eq. (\ref{diffeqpol}). Note that $z^T_B$
does not depend on $R$ and that similarly $V^T_B$ does not depend on
$y$. This fact is nevertheless an artefact of the approximation we
made. To study its limits of validity, we can look at (\ref{zhigh}) and (\ref{Vhigh}) and
require that the corrections are small: From this we can infer the
conditions 
\bea 
y^2\ll {1\over \beta \ln
\beta}\,,\,\,\,\,\,\,\,~~~~~~~R^2\ll {1\over N\beta^2 \ln \beta}. 
\label{condi}
\eea

We can now find the metric at first order in the low $\beta$
expansion, i.e. $z^T_B=1/2-N\beta$ and $V^T_B=0$. The metric in the LLM
coordinates is quickly computed and reads 
\bea 
ds^2= {y\over
\sqrt{N\beta}}\left(-dt^2+d\Omega_3^2\right)+{\sqrt{N\beta}\over y
}\left(dy^2+y^2d\tilde\Omega_3^2+dR^2+R^2
d\phi^2\right).\label{first} 
\eea 
Rescaling the coordinates as 
\bea
\tilde t=(N\beta)^{-1/4} t, \qquad \tilde y=(N\beta)^{1/4}
y, \qquad \tilde R=(N\beta)^{1/4} R 
\eea 
brings the metric into the form
\bea 
ds^2= (N\beta)^{1/4}\tilde y\left(-d\tilde t^2+{1\over
\sqrt{N\beta}} d\Omega_3^2\right)+ {1\over (N\beta)^{1/4}\tilde y}
\left(d \tilde y^2+\tilde y^2d\tilde\Omega_3^2+d\tilde R^2+\tilde
R^2 d\phi^2\right)\label{high}. 
\eea
This form of the metric closely resembles the dilute gas approximation limit studied in \cite{Lin:2004nb}. 
There, one considers a configuration of droplets with area $A_i$ in the $(x_1,x_2)$ plane, and send the distance 
between the droplets to infinity by the rescaling  
\bea
x \rightarrow \lambda \tilde x, \qquad 
x' \rightarrow \lambda \tilde x, \qquad y \rightarrow \lambda \tilde y, \qquad \lambda \rightarrow \infty 
\eea
while keeping the droplets areas $A_i$ fixed. The corresponding metric reads
\bea
ds^2=H^{-1/2}\left[-d\tilde t^2+\lambda^2
d \Omega_3^2\right]+ H^{1/2}\left[d\tilde y^2+\tilde y^2
d\tilde\Omega^2_3+dx^idx^i\right]
\label{dilute}
\eea 
where the harmonic function $H$ is  
\bea 
H={1\over \pi}\sum_i{A_i\over
[(\tilde x-\tilde x_i^{'})^2+\tilde y^2]^2}. 
\label{H} 
\eea
Thus the metric eq. (\ref{dilute}) can be viewed as a multi-center solution for a 
stack of separated $D3$-branes, and corresponds to the $SO(4)$ invariant sector of the 
Coulomb branch of the gauge theory.

Upon the identification $\lambda=(N\beta)^{-1/4}$, 
one can see that the dilute gas limit $\lambda\rightarrow \infty$ is similar to the high temperature regime $\beta \rightarrow 0$ of the thermal solution eq. (\ref{high}). This is perhaps not surprising since 
in the high temperature limit the fermion density goes to zero. We also notice that a 
continuum version of eq. (\ref{H}) with $A_i \equiv d^2 \tilde{x}' / \sqrt{N\beta}$ gives 
\bea 
H={1\over \pi} \int {d^2 \tilde{x}' \over \sqrt{N\beta}} \,{1 \over [
(\tilde{x}-\tilde{x}')^2+\tilde{y}^2]^2}={1\over \sqrt{N\beta} \, \tilde{y}^2}
\eea
which is what we would expect in order to match eq. (\ref{high}) with eq. (\ref{dilute}). 

Taking into account the next to leading order corrections for $z^T_B$
and $V^T_B$ in eq. (\ref{zhigh}) and (\ref{Vhigh}), we obtain the
metric 
\bea 
ds^2 &= & H^{-1/2}\left[-dt^2+
(1-N\beta)d\tilde\Omega_3^2\right]+H^{1/2}\left[d y^2+(1+N\beta)
y^2 d\Omega^2_3+dR^2+R^2 d\phi^2\right]\nonumber\\ 
&&+\sqrt N R^2 y\beta^{3/2}\ln\beta dt d\phi 
\label{second}
\eea 
where 
\bea
H = {N\beta\over y^2}-{N^2\beta^2\over y^2}+{1\over 2}\beta^2
N\ln\beta. 
\eea 
At this order we have a non vanishing $V^T_B$ and this
determines the presence of the mixed term $g_{\phi t}$ in the
metric.

%%%%%%%%%%%%%%%%%%%%%%%%%%%%%

\subsection{Energy and angular momentum}

We remark that the region of validity of the approximations made
so far does not allow us to use the metrics (\ref{first}) and
(\ref{second}) in the asymptotic region $R^2+y^2\gg 1$, because of the conditions 
eq. (\ref{condi}). Therefore,
to compute the energy of the hyperstar in the high temperature
regime, we need to find the form of the metric in the
complementary region of validity. The new metric will be trustable
in the asymptotic region and will allow a calculation of the
energy with the methods already discussed. To this end, it is convenient to first introduce  
polar coordinates in the $(x_1,x_2,y)$ space
\bea
R &=& u \cos \vartheta \cr
y &=& u \sin \vartheta.
\eea
Then one can evaluate eq. (\ref{zTB}) and eq. (\ref{VTB}) in an expansion for $u \gg 1$  
while keeping $T$ fixed but large (such that we are in the Boltzmann regime). The integrals 
involved in the expansion can be readily computed analytically and one ends up with the 
result
\bea
z^T_B (u,\vartheta) &=& {1 \over 2} -2 N \sin^2 \vartheta {1 \over u^2} - 8 \, NT \sin^2 \vartheta (3 \cos^2 \vartheta-1) {1 \over u^4} \cr 
&& -48 \, NT^2 \sin^2 \vartheta (10 \cos^4 \vartheta -8 \cos^2 \vartheta +1) {1 \over u^6} 
+ {\cal O} \left({1 \over u^8}\right) 
\eea   
\bea
V^T_B (u,\vartheta) &=& 2 N \cos^2 \vartheta {1 \over u^2} + 8 \, NT \cos^2 \vartheta (3 \cos^2 \vartheta-2) {1 \over u^4} \cr 
&& + 48 \, NT^2 \cos^2 \vartheta (10 \cos^4 \vartheta -12 \cos^2 \vartheta +3) {1 \over u^6} 
+ {\cal O} \left({1 \over u^8}\right)
\eea
where in the expansion we have kept only terms which contribute to the mass and angular momentum. 
One can now go to the $\as$ coordinates via the change of variables given in eq. (\ref{adscoord}) and use $z^T_B$ and $V^T_B$ to obtain the asymptotic form of the metric. The explicit expressions 
are somewhat lengthy and we will not report them here in detail. The computation of $M$ and $J$ follows the same lines of the one given in detail for the low temperature regime. Particularly straightforward is the evaluation of the angular momentum, which can be read off from the shift vector ${\cal N}^{\tilde \phi}$. The explicit calculation  gives
\bea
{\cal N}^{\tilde \phi} &=&  \left ({4 T \over L^3} - L \right) {1 \over r^2} + {\cal O} \left({1 \over r^4}\right) \cr
&=& {2 L \over N^2 r^2} \left( NT-{N^2 \over 2} \right) + {\cal O} \left({1 \over r^4}\right)
\eea 
where we have used the relation $N=L^4/2$ which holds in our units. Viewing ${\cal N}^{\tilde \phi}dt \equiv A$ as a gauge field in 5 dimensions, the angular momentum 
is equal to the corresponding electric charge, as explained in the previous section. The result is then
\bea
J = NT-{N^2 \over 2}. 
\eea 
This is indeed what we would have expected, since $NT$ is the energy for a gas of $N$ particles with Boltzmann 
density and $N^2 /2$ is the ground state energy of the $N$ fermions.

To compute the mass, we used both methods described in sec. 4. Once again, quadratic terms 
in the charge $(NT-N^2/2)$ appear in the calculation, with different coefficients in the two methods. The linear term 
is however scheme independent and gives the correct result
\bea
M = \left (NT-{N^2 \over 2} \right ) / L.
\eea

%%%%%%%%%%%%%%%%%%%%%%%%%%%%%%%%%%%%%%
\section{Conclusion and open questions}
%%%%%%%%%%%%%%%%%%%%%%%%%%%%%%%%%%%%%

In this paper we explored the thermodynamic properties of a 1/2
BPS IIB supergravity solution called hyperstar. This background was
first obtained in \cite{Buchel:2004mc} by thermal coarse-graining of
the ``bubbling AdS geometry" found in \cite{Lin:2004nb}. The
hyperstar is in correspondence with a distribution of free fermions
in thermodynamic equilibrium at temperature $T$, living on a two-dimensional phase space contained in the ten-dimensional geometry.

We studied both limits of low and high temperature. In the former
case, the fermions obey the Fermi-Dirac distribution and the
supergravity background is obtained from the LLM Ansatz by means of
a Sommerfeld expansion. We found agreement between the energy of the
fermions and the ADM mass of the supergravity,
modulo a subtlety involving $T^4$ terms which we discussed in the
main text. We also proposed a way to match the entropy of the fermions with the
entropy of the hyperstar in the low temperature limit. String $\alpha'$ corrections are expected to generate a finite area stretched horizon, lifting the naked singularity of the
hyperstar to a true black hole singularity.

In the classical limit of high temperature, we found the explicit
form of the metric of the supergravity background and we observed
how this metric resembles the metric of a dilute gas of D3 branes,
which corresponds to the $SO(4)$ invariant sector of the  Coulomb branch
of the CFT. We also computed the associated mass and angular momentum. 

It would be interesting to push this study further. 
An important point, as already remarked, would be to understand better the meaning
of the temperature for the supergravity solution. On a more fundamental level, it is worthwile to understand the exact relation
between a thermalized solution like the hyperstar and the Matrix Model description of the half-BPS sector of the dual CFT, extending considerations already made in \cite{Balasubramanian:2005mg}. 

Another issue  is whether the appearance of the naked singularity
in the hyperstar can be understood in terms of a distribution of giant gravitons, as is the case for the superstar \cite{Myers:2001aq}. 

The LLM geometries, upon dimensional 
reduction to five dimension, can be
seen as interesting generalizations of AdS half-BPS extremal black holes
 \cite{Chong:2004ce}.
It would then be interesting to obtain the explicit dimensional 
reduction to five dimension of the hyperstar. In this setting one could use the powerful methods of holographic renormalization to carry out the computation of the ADM mass. Then one could prove in a rigorous way that the quadratic contributions to the mass are effectively spurious and can be eliminated within an appropriate renormalization scheme.  

Finally, we would like to mention that 
the LLM construction has been extended to other BPS sectors of type IIB supergravity, see, for instance, \cite{Liu:2004ru} for the 1/4 BPS sector. In this case, one modifies
the LLM Ansatz in order to accomodate an axion-dilaton field which
breaks the supersymmetry by half. The effect of this field is to
introduce a deficit angle in the ``phase space". 
One could try to understand whether this phase space can be useful to study
the mass and entropy of the corresponding supergravity geometry.

%%%%%%%%%%%%%%%%%%%%%%
\vskip 15mm
\subsubsection*{Acknowledgments}

It is a pleasure to thank Gary Gibbons and Martin Ro\v cek for discussions
and Steven Gubser, Gary Horowitz, Hong Lu, Oleg Lunin and Simon Ross
for helpful correspondence. We acknowledge partial financial support
through NSF award PHY-0354776. 

%%%%%%%%%%%%%%%%%%%%%%%

%%%%%%%%%%%%%%%%%%%%%%%%%%%%

\end{document}